\documentclass[english,prl,twocolumn]{revtex4}
\usepackage[T1]{fontenc}
\usepackage[latin1]{inputenc}
\usepackage{graphicx}
\usepackage{amsmath}
\usepackage{aas_macros}
\usepackage{color}

\makeatletter
\usepackage{babel}
\makeatother
\begin{document}

\title{Imprints of spherical non-trivial topologies on the CMB}

\author{Anastasia Niarchou, Andrew Jaffe}

\affiliation{Imperial College London}

\date{\today{}}

\begin{abstract}
The apparent low power in the CMB temperature anisotropy power spectrum derived from WMAP motivated us to consider the possibility of a non-trivial topology. We focus on simple spherical multi-connected manifolds and discuss their implications for the CMB in terms of the power spectrum, maps and the correlation matrix. We perform Bayesian model comparison against the fiducial best-fit $\Lambda$CDM based both on the power spectrum and the correlation matrix to assess their statistical significance. We find that the first year power spectrum shows a slight preference for the Truncated Cube space, but the 3-year data show no evidence for any of these spaces.
\end{abstract}


\maketitle

\section{Introduction}
\label{sec:intro}

The advent of high-precision cosmological observations has revived interest in the topology of the Universe. The latest measurements leading to questioning our assumption of a simply-connected topology came from WMAP \cite{wmap1basic, wmap1ps}.
The anomalously low power at large scales in the temperature power spectrum, $C_\ell$, and suggestions of preferred directions \cite{wmap1ps, alignment} in the first year data could not be accounted for in the context of the standard flat $\Lambda$CDM model. A non-trivial topology could be a possible explanation. The observed low power, also reflected in the temperature correlation function, would arise naturally in this context as a consequence of the finite volume of the Universe. Thus, a number of (mainly flat), non-trivial topologies have been invoked, and  various statistical tests employed to infer their validity (e.g. \cite{bond, circles, hajian, phillips, kunz, circles1, circles2}). We study some of the simplest spherical non-trivial topologies. A closed geometry is marginally consistent with the data ($\Omega_k \equiv 1-\Omega_{0}=-0.02\pm0.02$). Moreover, the size of spherical manifolds (i.e., the length in various directions) is fixed with respect to the curvature radius, leaving only the curvature as a free parameter.
Spherical manifolds have been discussed before, mainly with regards to their power spectra \cite{spher1, spher2, spher3}. Here, we expand upon this work, presenting full correlation matrices and maps and use Bayesian model comparison (first applied in cosmological model selection in \cite{h0odds}) to assess their viability. 

We select our models using the geometrical degeneracy \cite{EfstBondConfusion99}, which states that universes of different spatial curvature have identical $C_\ell$ at small scales if their primordial power spectra, matter density and acoustic peak location parameter are the same. The degeneracy breaks at the large scales due to the ISW effect. We used this degeneracy in \cite{mypaper} to construct $C_\ell$ of simply-connected closed models showing reduced power at low $\ell$. We thus ensure our models exhibit the same high-$\ell$ $C_\ell$ as the WMAP best fit and focus on the low-$\ell$ regime, where topology plays an important role (we examine the $2\le\ell\le10$ region). The degeneracy constrains all cosmological parameters but the curvature, which is therefore related to the Hubble parameter; the two can be used as free parameters interchangeably \cite{mypaper}. We explore $-0.063\le\Omega_k\le-0.008$ corresponding to $52\le H_0\le68$ km/sec/Mpc.

\section{Spherical Non-trivial Topologies And Their Imprints}

We focus on the Quaternionic, Octahedral, Truncated Cube and Poincar\'{e} spaces. Their topological properties are listed in \cite{gaussmann, ghost}. Their fundamental domains are a four-sided prism, a regular octahedron, a truncated cube and a regular dodecahedron and the number of fundamental domains tesselating the 3-sphere is 8, 24, 48 and 120 respectively. These manifolds are rigid, indicating they have no other degrees of freedom aside from their orientation with respect to the coordinate system. The CMB temperature anisotropy spherical harmonic expansion coefficients and correlation matrix $C_{{\ell}m}^{{\ell}'m'}\equiv\langle a_{\ell m}a^*_{\ell'm'}\rangle$ take the form
\begin{eqnarray}
\label{eq:almmc}
a_{{\ell}m}&=&
\sum_{\beta}\Delta_{\ell}(\beta)\sqrt{P(\beta)}\sum_s{\xi}_{\beta{\ell}m}^s\hat{e}_{{\beta}s}\nonumber\\
C_{{\ell}m}^{{\ell}'m'}&=&
\sum_{\beta,s}\Delta_{\ell}(\beta)\Delta_{\ell'}(\beta)P(\beta){\xi}_{\beta{\ell}m}^s{\xi}_{\beta{\ell}'m'}^{s*}\nonumber
\end{eqnarray}
where 
$\Delta_{\ell}$ are the CMB transfer functions extracted from CMBFAST, $P(\beta)$ is the primordial power spectrum of perturbations and the $\hat{e}$ are Gaussian random numbers. The $\xi_{\beta\ell m}^s$ are particular to the space in question and encode its topological properties. 
They are computed via a variant of the `ghost method' \cite{ghost}, which we will describe in detail in \cite{preparation}. The $\beta=(k+1)\sqrt{K}$ are discrete wavenumbers where $k>1$ is an integer and $K$ is the (unnormalized) curvature. Finally, we note that, while in a simply-connected trivial topology all $k\ge2$ correspond to eigenvalues of the Laplacian, this is not the case with the multi-connected topologies  \cite{ghost}. In our analysis, we used $\beta\le41$ for $2\le\ell\le10$.

\begin{figure}[htbp]
\centering\includegraphics[width=0.99\columnwidth]{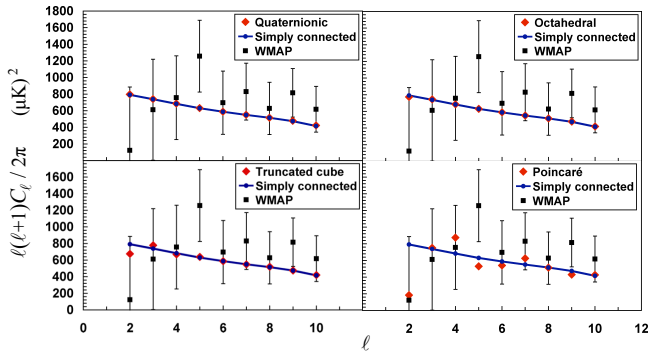}
\caption{Power spectra of simply-/multi-connected spherical spaces for $\Omega_k=-0.017$ and a constant spectral index and from the first-year WMAP data. The drop at higher $\ell$ is partly artificial and due to the small number of wavenumbers we used ($\beta\le41$).}
\label{fig:psp64}
\end{figure}

Spherical multi-connected spaces exhibit reduced mode density (which can result in suppression of power) and anisotropic correlations at large scales. Fig.~\ref{fig:psp64} shows the $C_{\ell}$ of the four spaces of interest for $\Omega_k=-0.017$.

A decrease in the size of the fundamental domain and the number of allowed eigenmodes increases the suppression at large scales and the power spectrum appears more jagged. These effects become stronger if the curvature increases and at very high $\Omega_k$ the Octahedral space shows a power deficit as well, supporting the argument \cite{spher1} that small well-proportioned universes suppress the quadrupole. The fit of the $C_\ell$ of the Truncated Cube and Poincar\'{e} spaces is worth noting; however, it requires fine-tuning $\Omega_k$.

\begin{figure}[htbp]
\centering\includegraphics[width=0.99\columnwidth]{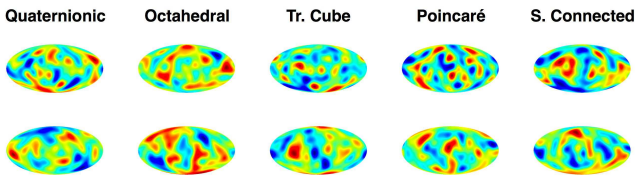}
\caption{Maps for spherical non-trivial/trivial spaces. Top row: $\Omega_k=-0.063$. Bottom row: $\Omega_k=-0.017$.}
\label{fig:maps}
\end{figure}

Our manifolds are intrinsically anisotropic, inducing apparent non-Gaussianity on CMB maps (i.e., if analyzed assuming isotropy the distribution of multipole moments at a single $\ell$ might seem inconsistent with a univariate Gaussian). We constructed maps for $2\le\ell\le10$ and show a sample in Fig.~\ref{fig:maps}. There appears to be some structure in the maps of the two smallest spaces when $\Omega_k=-0.063$, but when the curvature decreases all maps appear interchangeable. Of course, we would have to analyze a large sample to identify any patterns. Upon decomposition of our realisations into individual multipoles, we found occurrences of alignment; however, we need more realisations to draw any definite conclusions and the alignments we found may be isolated, pathological features.

\begin{figure}[htbp]
\centering\includegraphics[width=0.99\columnwidth]{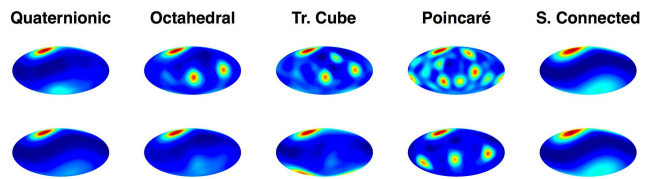}
\caption{A row of $C_{{\ell}m}^{{\ell}'m'}$ for spherical non-trivial/trivial spaces showing correlations with the point in the upper left. Top row: $\Omega_k=-0.063$. Bottom row: $\Omega_k=-0.017$.}
\label{fig:pixcorr}
\end{figure}

Signatures of multi-connectedness are much more apparent in the correlation matrix. Fig.~\ref{fig:pixcorr} shows visualisations of a row of $C_{{\ell}m}^{{\ell}'m'}$ in pixel space. The Quaternionic space shows no structure, the Octahedral space only shows correlations when $\Omega_0$ is high and the other two spaces always show correlations. The extent of these patterns grows as the fundamental domain becomes smaller and the topology more complex. These correlations appear when the distance to the LSS, $\chi_{LSS}$, becomes larger than the injectivity radius $R_I$ (half the smallest geodesic between an object and its image) of a given topology. In our models, $\chi_{LSS}$ varies (in units of the curvature radius) from 0.649 (for $\Omega_k=-0.063$) to 0.280 (for $\Omega_k=-0.007$). The $R_I$ for the Quaternionic, Octahedral, Truncated Cube and Poincar\'{e} spaces are 0.785, 0.524, 0.397 and 0.314 respectively \cite{radii}. In our case, for the Quaternionic space $R_I>\chi_{LSS}$ always, and for the Octahedral, Truncated Cube and Poincar\'{e} spaces $R_I>\chi_{LSS}$ at $\Omega_k\sim-0.029, -0.012, -0.009$ respectively. Beyond these limits, it is impossible to identify any of these topologies using the circles-in-the-sky method; however, they can be detected by our analysis, since off-diagonal terms still exist in the correlation matrix.

We also experimented with varying the spectral index $n$ of the primordial power spectrum of perturbations: for a range of $n$ from 0.5 to 1.2 and various values of $H_{0}$ we found that a change in $n$ tilts the power spectrum as expected, but leaves the patterns in the correlation matrix largely unchanged.

\section{Model Comparison}

The suppression of power at low $\ell$ suggests that these multi-connected manifolds could explain the WMAP findings. In order to assess their statistical significance, we perform Bayesian model comparison as in \cite{mypaper}.

\paragraph{Power Spectrum.}

We first compare the non-trivial topologies to the fiducial best fit model with respect to the $2\le\ell\le10$ part of the power spectrum. We use both the first- and three-year power spectra, because they differ primarily as a result of a different analysis technique used by the WMAP team\cite{wmap3-powersp}. Thus, we will be able to identify the impact of using different data analysis techniques on the evidence for spherical topologies.

\begin{figure}[htbp]
\centering\includegraphics[width=\columnwidth]{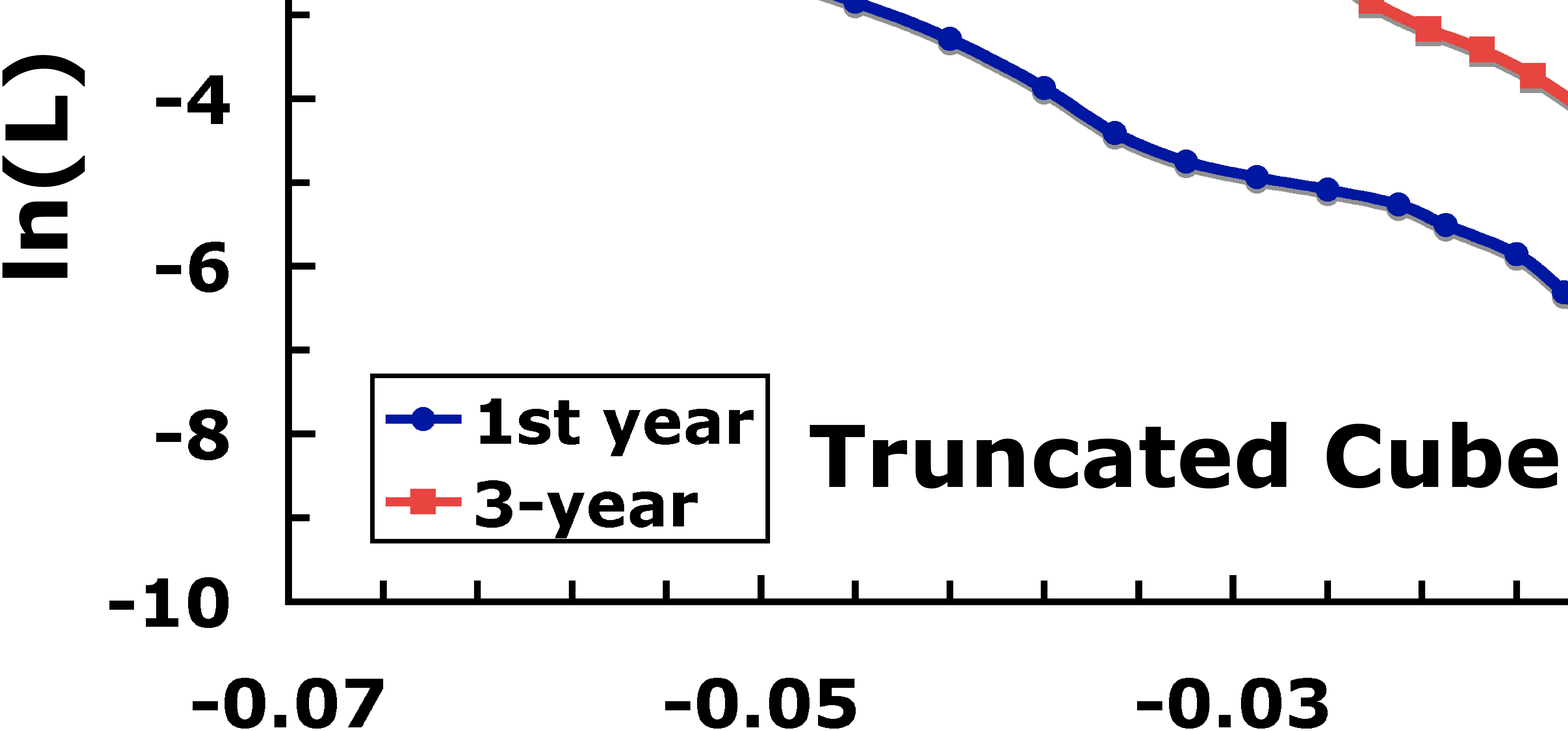}
\caption{Likelihood based on the power spectrum.}
\label{fig:powerlik}
\end{figure}

We express the likelihood as a function of the curvature and show it for both the first- and three-year WMAP data in Fig.~\ref{fig:powerlik}. In each case we have used the appropriate likelihood code offered by the WMAP team (note that for the three-year data we use the low-$\ell$ component of the WMAP likelihood code). 

We see that the first year data strongly favours a higher curvature (corresponding to lower values of $H_0$) and a smaller fundamental domain, as a smaller universe would show more suppression of power at low $\ell$, in agreement with the data. In the case of the Poincar\'{e} space, however, the highly-closed regime $\Omega_k\lesssim -0.06 $ regime is not favoured, because the $C_{\ell}$ do not fit the data well. The same trend is observed in the likelihood based on the three-year data. We note that the drop in the likelihood at higher curvature is partly artificial and due to the fact that we only used $k\le40$ in our calculations.

We present the evidence for each topology in Table \ref{tab:psp1}. We impose a flat prior on $\Omega_k$ in the range [$-0.063$, $-0.008$] (52  km/sec/Mpc $\le H_0 \le 68$ km/sec/Mpc). The upper bound is a result of computational limitations preventing us from getting accurate power spectra and correlation matrices at higher $\Omega_0$. The lower bound is based on numerous suggestions that the Hubble constant is not likely to be any lower (see \cite{hubble} and references therein). The Octahedral and the Truncated Cube spaces are preferred by the first year data with Bayes factors of 2 and 8 respectively. The latter figure translates into definite, though not strong, evidence in favour of the Truncated Cube. However, the three-year data exclude all non-trivial topologies.

\begin{table}\centering\begin{tabular}{|c||c|c|c|} \hline
                        \bfseries{model} & \bfseries{Bayes factor} & \bfseries{$\sigma$} & \bfseries{preference}\\ \hline
                        best-fit & 1\,\; \vline\ {\bfseries{1}} & -- & --\\ \hline
                        Quaternionic & 0.864\,\, \vline\ 0.077 & 0.54 \vline\ 2.26 & \it{best-fit}\\ \hline
                        Octahedral & 2.031\,\, \vline\ 0.320 & 1.19 \vline\ 1.51 & \it{Octahedral \vline\ best-fit}\\ \hline
                        Tr. Cube & {\bfseries{8.064}} \vline\ 0.142 & 2.04 \vline\ 1.98 & \it{Tr. Cube\,\,\,\, \vline\ best-fit}\\ \hline
                        Poincar\'{e} & 0.400\,\, \vline\ 0.037 & 1.35 \vline\ 2.57 & \it{best-fit}\\ \hline\hline
best-fit&{\bfseries{1}}&--&--\\ \hline
Quaternionic & 0.0395 & 2.54 &\it{best-fit}\\ \hline
Octahedral & 0.0047 & 3.28 &\it{best-fit}\\ \hline
Tr. Cube & 0.0003 & 4.01 &\it{best-fit}\\ \hline
Poincar\'{e} & $\ll 1$ & $\gg 1$ &\it{best-fit}\\ \hline
                       \end{tabular}

\caption{\label{tab:psp1} {\it{Top panel}}: Model comparison using $C _\ell$ for the first (1st col.) and three-year (2nd col.) WMAP data. {\it{Bottom panel}}: Model comparison using the correlation matrix. The Bayes factor of the preferred model is shown in bold.}
\end{table}

If we switch from the agnostic flat prior on $\Omega_k$ to a Gaussian with $H_0=72\pm10\;\textrm{km/sec/Mpc}$ (allowing for systematic error), we find a drop in the evidence for non-trivial topologies based on the first year data. This is expected, since this prior emphasizes higher values of $H_0$, where the likelihood is lower. Now only the Truncated Cube space is preferred, by a Bayes factor of just 2. When taking into account the three-year data a change in the prior does not have a significant effect.

\paragraph{Correlation Matrix.}

Evidently, $\Omega_k$ is still a free parameter of our likelihood function; now, however, we must take into account the fact that $C_{{\ell}m}^{{\ell}'m'}$ is not rotationally invariant. We can express the dependence on orientation through Euler angles $\alpha, \beta, \gamma$ \cite{Varshalovich} with $0\leq\alpha\leq2\pi$, $0\leq\beta\leq\pi$ and $0\leq\gamma\leq2\pi$, realised by the Wigner D-functions \cite{Edmonds, Varshalovich}. We have also let the amplitude of the power spectrum of initial perturbations  vary slightly.
Now the likelihood takes the form:
\begin{equation*}
\label{eq:lik3}
{\cal{L}}(A,\Omega_{k},\alpha,\beta,\gamma)=\frac{1}{\sqrt{|2\pi C_{A,\Omega_{k}}|}}\exp-\frac12{\left[Ra\right]^{T}C^{-1}_{A,\Omega_{k}}\left[Ra\right]},
\end{equation*}
where the $A, \Omega_{k}$ subscript denotes that the correlation matrix depends on the curvature and the amplitude of initial perturbations and $R\equiv R(\alpha,\beta,\gamma)$ is the rotation matrix (Wigner function).
The amplitude $A$ \cite{wmap1basic} (the normalisation factor in CMBFAST with the option ``unnorm'') is related to the amplitude of fluctuations at horizon crossing via $|\Delta_R(k_0)|^2=2.95\times10^{-9}A$. The data we use are the $a_{\ell m}$s extracted from the three-year WMAP ILC map whose noise at large scales is negligible.

\begin{figure}[htbp]
\centering\includegraphics[width=0.99\columnwidth]{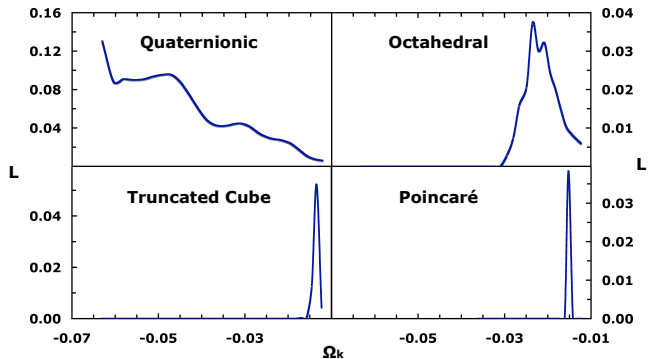}
\caption{Likelihood based on the correlation matrix. The likelihood of the Poincar\'{e} space is multiplied by $10^8$.}
\label{fig:corrlik}
\end{figure}

Marginalizing the likelihood function over the amplitude (with a flat prior in the range [0.5, 1.2]) and the Euler angles (we will explore the structure of the likelihood in more detail in \cite{preparation}), shows that the data favours spaces with little or no structure in their correlation matrices (Fig.~\ref{fig:corrlik}). Indeed, the likelihood function takes higher values for less complex spaces, where correlations among various multipoles are weak. 

Finally, we calculate the Bayes factor for each topology using the same prior on $\Omega_k$ (Table \ref{tab:psp1}).
A comparison between the two panels shows that the inclusion of off-diagonal terms decreases the chances of all non-trivial topologies but the Quaternionic space. However, its Bayes factor is still too small to grant it any preference over the standard  model. Changing to a Gaussian prior on $H_0$ does not affect these results significantly. A computation of the evidence for the Truncated Cube space with $n=0.5$ and $n=1.2$ yields Bayes factors of $10^{-8}$ and $2.6\cdot10^{-5}$ respectively. Given the effect of changing $n$, we expect this result to reflect mostly the change in the averaged diagonal power spectrum.

\section{Conclusion}

The non-trivial topologies we have examined are not a viable candidate for explaining the low-$\ell$ anomaly in the WMAP data. Bayesian model comparison shows that the flat, simply-connected $\Lambda$CDM model fits the data better. Considering only the power spectrum (assuming a constant $n$ over our range of $\beta$) topological spaces are penalized for requiring a low Hubble constant and tuning of the curvature, while a full analysis shows that the observations do not support the extent of structure present in their correlation matrices. 

However, the possibility of a multi-connected topology is not necessarily excluded. We have limited ourselves to the simplest spherical multi-connected manifolds; double- or linked-action manifolds have not been explored (however calculating their Laplacian eigenmodes could be computationally challenging). More importantly, our models required a low Hubble constant as a result of the geometrical degeneracy. It might be possible to adjust the power at large scales using other combinations of cosmological parameters, but this would require an additional mechanism to ensure that the shape of the power spectrum at small scales matches the observations. Finally, it is possible that the primordial power spectrum is actually different from what we used. Lacking an established mechanism generating perturbations in closed spaces, we adopted the analogue of a scale-invariant power spectrum.  If such a mechanism is ever conceived, it might predict a different form of a primordial power spectrum (although a change in the spectral index does not affect the pattern of correlations much). Another possibility would be to search for more information in other aspects of CMB observations, such as polarisation, where a circles-in-the-sky approach might yield some results \cite{riazuelo4}. But even in this case, a full analysis would have to follow the steps we described above.

In conclusion, although the current data and the cosmological theories at hand do not support the case for a non-trivial topology, the number of possibilities left to explore suggests that the issue of the topology of the Universe is far from settled.
\\

\small{The authors wish to thank J. Weeks for useful information and an implementation of the original `ghost method', and the anonymous referees for useful suggestions.}

\bibliography{topology}

\end{document}